\documentclass[aps,prl,reprint,twocolumn,oneside,floatfix,sort&compress,
superscriptaddress,showpacs,nobalancelastpage]{revtex4-1}

\usepackage{graphicx,xcolor}
\usepackage[utf8]{inputenc}
\usepackage{amssymb,bbm}
\usepackage[reqno]{amsmath}

\usepackage[colorlinks,bookmarks=false,citecolor=blue,linkcolor=red,urlcolor=blue,hyperfootnotes=true]{hyperref}

\begin{document}

\title{Matrix product state formulation of frequency-space dynamics at finite temperatures}

\author{Alexander C. Tiegel}
\affiliation{Institut f\"ur Theoretische Physik, Georg-August-Universit\"at G\"ottingen, 37077 G\"ottingen, Germany}

\author{Salvatore R. Manmana}
\affiliation{Institut f\"ur Theoretische Physik, Georg-August-Universit\"at G\"ottingen, 37077 G\"ottingen, Germany}

\author{Thomas Pruschke}
\affiliation{Institut f\"ur Theoretische Physik, Georg-August-Universit\"at G\"ottingen, 37077 G\"ottingen, Germany}

\author{Andreas Honecker}
\affiliation{Institut f\"ur Theoretische Physik, Georg-August-Universit\"at G\"ottingen, 37077 G\"ottingen, Germany}
\affiliation{Fakult\"at f\"ur Mathematik und Informatik, Georg-August-Universit\"at G\"ottingen, 37073 G\"ottingen, Germany}

\date{\today}                 


\begin{abstract}
We present a flexible density-matrix renormalization group approach to calculate finite-temperature spectral functions of one-dimensional strongly correlated quantum systems. The method combines the purification of the finite-temperature density operator with a moment expansion of the Green's function. Using this approach, we study finite-temperature properties of dynamical spectral functions of spin-$1/2$ $XXZ$  chains with Dzyaloshinskii-Moriya interactions in magnetic fields and analyze the effect of these symmetry breaking interactions on the nature of the finite-temperature dynamic spin structure factor.
\end{abstract}

\maketitle


In the study of strongly correlated quantum systems momentum and frequency-resolved spectral functions are of great interest as they provide important insights into the governing many-body physics \cite{FetterWalecka_1971}. For instance, experimental advances in neutron scattering \cite{Zaliznyak_2005} or electron spin resonance \cite{Zvyagin_2012_Review} allow for a very precise measurement of dynamical response functions. The availability of efficient and accurate numerical tools is thus highly desirable for making theoretical predictions at finite temperatures $T>0$.  Moreover, tuning the temperature can lead to interesting phenomena, such as the transition to a spin incoherent Luttinger liquid in one dimension \cite{Fiete_RMP}, or probing the quantum critical regime in correlated materials~\cite{Sachdev}. 
Established numerical methods such as quantum Monte Carlo (QMC) \cite{Sandvik_review,Sandvik_1991, Grossjohann_2009} or exact diagonalization (ED)~\cite{Sandvik_review} are indispensable, but are restricted either by the fermionic sign problem and a challenging analytical continuation procedure or small system sizes, respectively. On the other hand, series expansion techniques are limited to high temperatures \cite{Starykh_1997}.

In one dimension, the density-matrix renormalization group (DMRG) \cite{White_1992,White_1993,dmrgbook,schollwoeck2005,Schollwoeck201196} arguably offers efficient approaches to spectral functions. At zero temperature, DMRG-based frequency-domain methods have fostered significant progress in the study of spectral functions \cite{Hallberg_1995, Ramasesha_1996, Kuehner_1999, Jeckelmann_2002, Weichselbaum_2009}. Most prominently, the dynamical DMRG \cite{Jeckelmann_2002} provided new insights, e.g. into the dynamical properties of charge transfer salts such as TTF-TCNQ \cite{Benthien_2004}. The calculation of spectral functions is also possible in the context of the time-dependent DMRG (t-DMRG)~\cite{Vidal_2004, White_2004, Daley_2004, Schmitteckert_2004}. This approach has been successfully extended to finite temperatures \cite{Verstraete_2004, Zwolak_2004, Feiguin_2005, Pizorn_2014}, where either a purification of the density matrix or a formulation in terms of matrix product operators is used to compute response functions at $T>0$ via a real-time evolution; alternatively, a combination of the finite-temperature Lanczos approach~\cite{JaklicPrelovsek_1994, JaklicPrelovsek_2000} with DMRG has been proposed which requires stochastic sampling~\cite{KokaljPrelovsek_2009}. 

Our starting point is the observation that the accuracy of the t-DMRG approach is restricted by the maximal accessible time \cite{Barthel_2009, Feiguin_2010}, which is due to the growth of entanglement in the course of the t-DMRG procedure. Recently, the entanglement growth has been reduced by time evolving the auxiliary degrees of freedom backward in time \cite{Karrasch_2012_PRL,Karrasch_2013}, and by further related optimization schemes \cite{Barthel_2013}. Nevertheless, in order to gain better access to the low-frequency properties, it would be highly desirable to work with methods not relying on the hardly accessible long-time behavior of response functions.

In this Rapid Communication, we present a DMRG-based finite-temperature approach working directly in the frequency domain. We do so by considering the Liouville space dynamics of the purified density matrix. As the purification is a pure-state wave function in a doubled Hilbert space, it can be associated with a Liouville space vector~\cite{Barnett_1987}. Hence, dynamical correlation functions at $T>0$ can be approximated by a moment expansion of the Green's function. Here we implement this idea in the framework of matrix product states (MPS) using an expansion in Chebyshev polynomials \cite{Weisse_2006, Holzner_2011} with respect to the Liouville operator.  

This allows us to study finite-temperature properties of low-dimensional systems with a higher frequency resolution than in previous developments relying on real-time evolution. Motivated by experiments on quasi-one-dimensional (1D) materials such as copper pyrimidine dinitrate (Cu-PM) \cite{Zvygain_2004, Zvyagin_2011} or copper benzoate \cite{Dender_1996, Asano_2003}, we apply this technique to study finite-temperature properties of dynamical spectral functions of spin-$1/2$ $XXZ$ chains in a uniform longitudinal magnetic field by adding a staggered transverse field mimicking the Dzyaloshinskii-Moriya (DM) interactions in these materials. 

\paragraph{Method.}
In order to extend the DMRG to $T>0$, one approach is to purify the density matrix $\rho_T$ \cite{Schollwoeck201196}. This is achieved by working in a doubled Hilbert space consisting of the physical state space $\mathcal{H}_P$ and an auxiliary space $\mathcal{H}_Q$ chosen to be isomorphic to $\mathcal{H}_P$. Then the pure state $| \Psi_T \rangle$ is an element of the tensor product space $\mathcal{H}_P \otimes \mathcal{H}_Q$, so that the density operator of the physical system is given by $\mathrm{Tr}_Q\,| \Psi_T \rangle \langle \Psi_T |$. As explained in detail in Ref.\ \onlinecite{Schollwoeck201196}, the desired thermal state is obtained via an imaginary time evolution starting at infinite temperature, $|\Psi_T \rangle =  e^{-(H_P \otimes I_Q)/(2T)}  |\Psi_{\infty} \rangle$, where $|\Psi_{\infty}\rangle$ denotes an initial state with maximal entanglement between the real and the auxiliary system. The Hamiltonian $H$ and the identity operator $I$ act on the spaces specified by the respective indices. The obtained thermal state $|\Psi_T \rangle$ of the doubled system corresponds to a vector in the Liouville space of operators \cite{Barnett_1987}. Thus, its dynamics is governed by the Liouville equation ($\hbar \equiv 1$ from now on)
\begin{align}
 \frac{\partial}{ \partial t} \,|\Psi_T \rangle   = - i \mathcal{L} \, |\Psi_T \rangle \label{eq: Liouville},
\end{align}
where $\mathcal{L} = H_P \otimes I_Q \,-\, I_P \otimes H_Q $ is the Liouville operator (see also
Ref.\ \onlinecite{Zwolak_2004} for a superoperator approach to mixed-state dynamics with MPS). Note that the backward time evolution on the auxiliary space, proposed to reduce the entanglement growth in the course of a real-time evolution \cite{Karrasch_2012_PRL}, can be motivated by the Liouvillian description. Equation~\eqref{eq: Liouville} can be solved via a Laplace transform in terms of the corresponding resolvent operator $G(z)=(z-\mathcal{L})^{-1}$ \cite{Dalton_1982}, where $z=\omega + i \eta$ is a complex frequency. This allows for the calculation of momentum and frequency-resolved dynamical response functions of the form 
\begin{align}
   I_A(k,\omega) = - \frac{1}{\pi} \, \text{Im} \, \left\langle \Psi_T \left| A^\dagger \frac{1}{z-\mathcal{L}} A \right| \Psi_T \right\rangle. \label{eq: specfunc}
\end{align}
Here $A = A_P \otimes I_Q$ denotes the observable of interest. The eigenvalues of the operator $\mathcal{L}$ are the differences of the eigenenergies of the Hamiltonian $H$. From this formulation it is evident that the computation of finite-temperature dynamics is inherently amenable to standard numerical methods working directly in the frequency domain. An approximation of the response function in Eq.\ \eqref{eq: specfunc} can, for instance, be obtained by a continued fraction expansion (CFE) \cite{Haydock_1972, Gagliano_1987, Hallberg_1995, Dargel_2011, Dargel_2012} or a Chebyshev expansion \cite{Weisse_2006, Holzner_2011, PhysRevB.90.045144}. For our proof-of-principle results, we use the latter because we found that an MPS-based expansion in Chebyshev polynomials has higher numerical stability and better convergence properties \cite{Weisse_epjb_2004}. The Supplemental Material features CFE results to show the flexibility of the Liouvillian formulation.

A Chebyshev expansion only grants convergence in the interval $\left[-1,1\right]$, since the Chebyshev polynomials $T_n(x)=\cos[n\arccos(x)]$ grow rapidly for $|x|>1$. Thus, we map the full many-body bandwidth $W$ of the Liouvillian $\mathcal{L}$ to $\left[-1,1\right]$, i.e.\ $\omega \in \left[-W/2, W/2 \right] \mapsto \omega' \in \left[-W', W'\right]$ according to $\omega' = (\omega+W/2)/a - W'$. The choice of $W'= 1- \epsilon/2$ with $\epsilon=0.025$ acts as a safeguard to strictly impose $\omega' \in \left[-1,1\right]$ and $a=W/(2 W')$. The rescaled Liouvillian is denoted by $\mathcal{L}' = (\mathcal{L}+W/2) /a - W'$. Instead of mapping the entire bandwidth to $\left[-1, 1\right]$, it should also be feasible to enhance the resolution by working with a smaller interval comparable to the width of the support of the spectral function \cite{Holzner_2011}.

Our finite-temperature DMRG calculations proceed as follows: First, we employ a Lanczos time evolution algorithm in MPS formulation~\cite{Park_1986, Noack_2005, Garcia-Ripoll_2006} for the imaginary time evolution to the desired thermal state $|\Psi_T \rangle$. The Chebyshev vectors $|t_n\rangle$, each represented as a pure state in the enlarged Hilbert space $\mathcal{H}_P \otimes \mathcal{H}_Q$, are generated via the recursion relation
\begin{align}
   | t_n \rangle = 2 \mathcal{L}'| t_{n-1} \rangle -  | t_{n-2} \rangle,
\end{align}
where $| t_0 \rangle = A| \Psi_T \rangle, \,\, | t_1 \rangle = \mathcal{L}'| t_0 \rangle$. 
With this notation, the expansion becomes
\begin{align}
 I_A(\omega) = \frac{2 W'/W }{\pi \sqrt{1- \omega'^2}} \left[ g_0 \langle t_0 | t_0 \rangle + 2 \sum_{n=1}^{N-1} g_n \langle t_0 | t_n \rangle T_n(\omega') \right].
\end{align}
The real numbers $g_n$ are damping factors which remove artificial oscillations occurring as consequence of the finite order $N$ of the expansion. We employ Jackson damping yielding a nearly Gaussian broadening $\eta(N)$ decreasing with $N$ \cite{Weisse_2006, Holzner_2011}. The computations are performed in real arithmetics, and we control the accuracy by specifying the dimension $m$ of the truncated Hilbert space. 

\begin{figure}[h]
\centering
 \includegraphics[width=0.89\columnwidth]{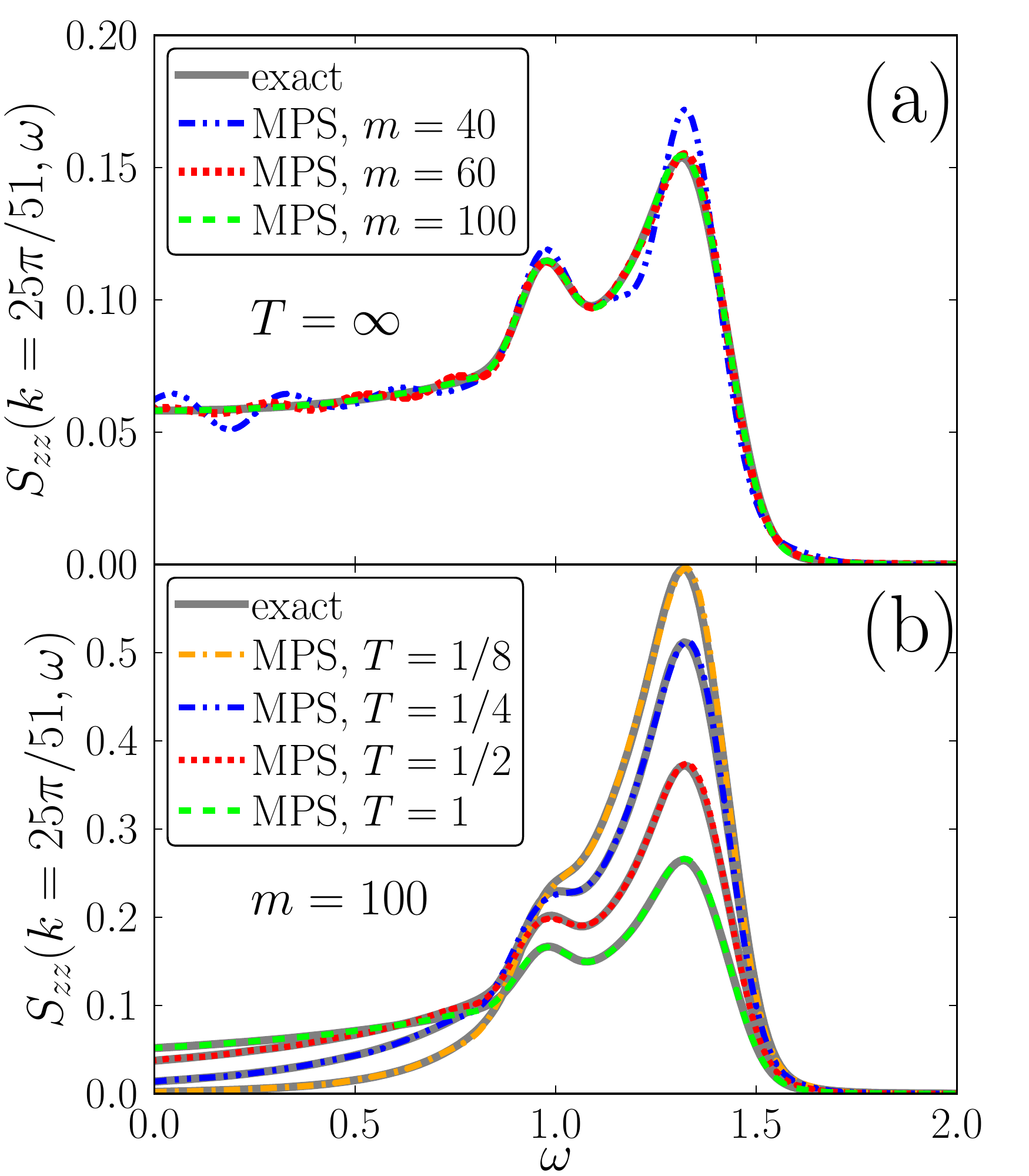} 
 \caption{(Color online) Finite-temperature DMRG calculations for the dynamic spin structure factor
 $S_{zz}(k=25\pi/51,\omega)$ of an open $XX$ chain with $L=50$ sites compared to exact results with Gaussian broadening $\eta=0.06$ ($N=1700$). (a) At  $T=\infty$ the accuracy is studied for different numbers of kept DMRG states $m$. (b) Temperature  dependence of $S_{zz}(k=25\pi/51,\omega)$ compared to the exact solutions (solid lines) for $T=0.125,\,\,0.25,\,0.5$, and~$1$ (from top to bottom).}
\label{fig: xy_chain_comparison}
\end{figure}

\paragraph{Comparison to exact results for the XX model.}

As a test case, we calculate the longitudinal spin structure factor  $S_{zz}(k,\omega)$ of the $XX$ model in zero field, 
\begin{align}
  H_{XX} = J \sum_{i=1}^{L-1}\, \left( S_i^x S_{i+1}^x + S_i^y S_{i+1}^y \right),
\end{align}
where $S^\alpha_i$ ($\alpha=x,y,z$) are the components of the spin operator $\mathbf{S}_i$ at site $i$, and we assume an antiferromagnetic exchange, $J \equiv 1$. 
By virtue of the Jordan-Wigner transform \cite{Lieb_1961}, this system is mapped to free fermions and is hence exactly solvable. We choose open boundary conditions (OBCs) as the standard DMRG is more efficient in this case \cite{Schollwoeck201196}. Following Ref.\ \onlinecite{Benthien_2004}, we define the spin operators in $k$ space as $ S^\alpha_k = \sqrt{\frac{2}{L+1}} \, \sum_{i=1}^L \, \sin(ki) \, S^\alpha_i $ with respect to the quasi momenta $k=\pi n / (L+1)$ and integers $n = 1,\ldots,L$. For the computation of
$S_{zz}(k,\omega)$, the operator of interest now is $A = (S^z_k)_P \otimes I_Q$. The time-dependent spin correlation functions $\langle S_i^z(t) S_j^z (0)\rangle$ can be evaluated exactly~\cite{Derzhko_1998, Derzhko_2000}. Fourier transforming the correlation functions  \footnote{More specifically, we formulate the Fourier transform into $k$- and $\omega$-space as \unexpanded{$S_{zz}(k,\omega) = \sum_{i,j} \frac{\sin(ki)\sin(kj)}{\pi(L+1)} \int \text{d}t \,e^{i\omega t - \eta t^2} \langle S_i^z(t) S_j^z(0)\rangle$}, where \unexpanded{$\langle S_i^z(-t) S_j^z(0) \rangle = \langle S_i^z(t) S_j^z(0)\rangle^*$} holds for negative times.},
we obtain the comparison to the MPS-based Chebyshev expansion of order $N=1700$ shown in Fig.~\ref{fig: xy_chain_comparison} for a system of size $L=50$. Figure~\ref{fig: xy_chain_comparison}(a) shows our results for the longitudinal spin structure factor
$S_{zz}(k=25\pi/51,\omega)$ at $T=\infty$ when varying the DMRG truncation $m$. Although the Gaussian broadening $\eta$ of the expansions is not strictly uniform by construction \footnote{For a $\delta$ function at $\omega'\in \left[-W',W' \right]$, Jackson damping gives a peak of width $\pi \sqrt{W'^2-\omega'^2}/N$.}, the agreement with the exact result ($\eta=0.06$) is excellent for $m=100$. In Fig.~\ref{fig: xy_chain_comparison}(b) the temperature dependence of $S_{zz}(k=25\pi/51,\omega)$ is depicted for $m=100$. Again, the MPS results fit the exact curves well down to temperatures as small as $T=0.125$. 

\begin{figure}[t] 
\centering
 \includegraphics[width=0.95\columnwidth]{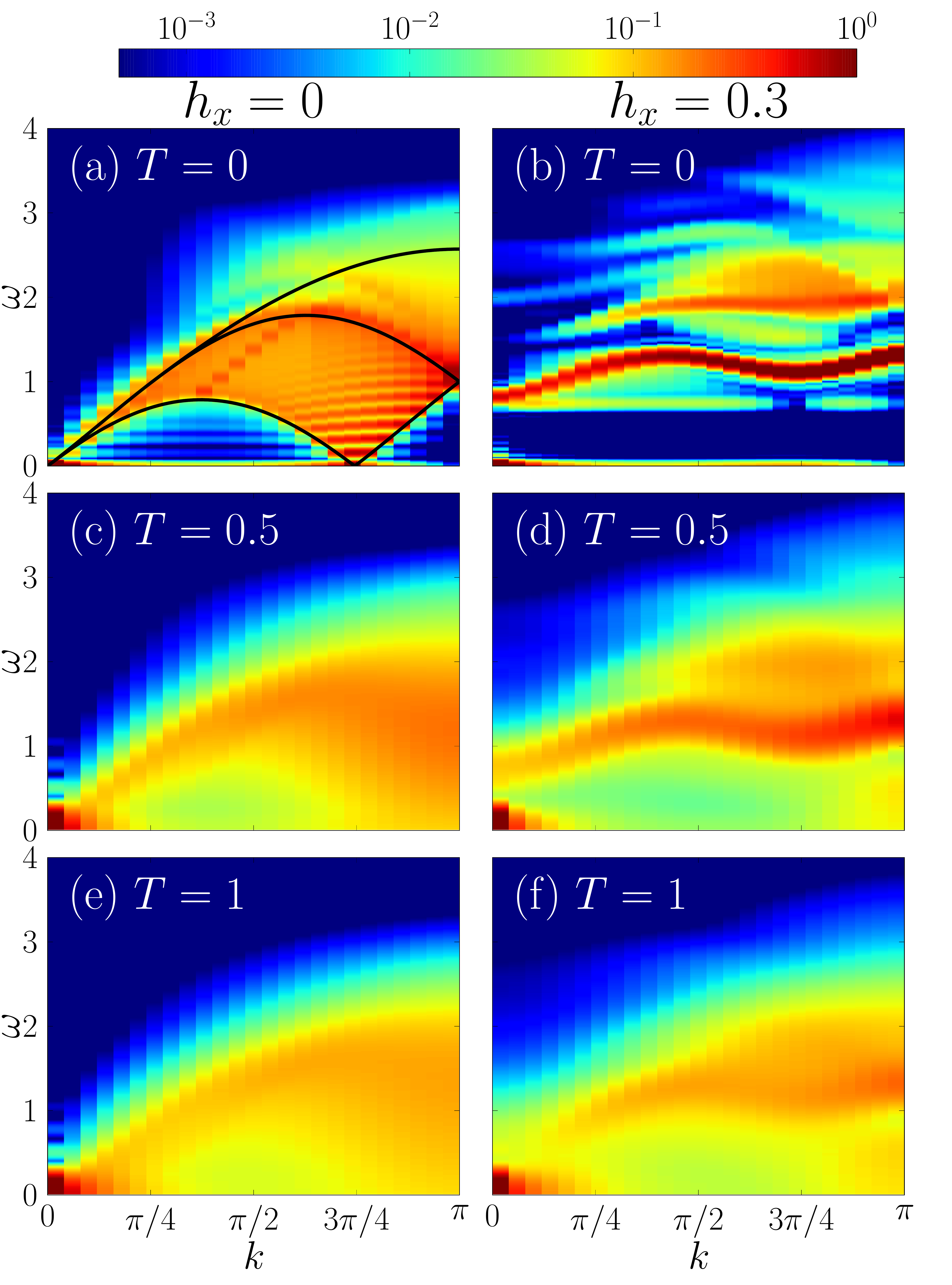} 
 \caption{(Color online) Longitudinal spin structure factor $S_{zz}(k,\omega)$ of a Heisenberg chain in a magnetic field $h_z = 1$ obtained by the Chebyshev MPS method with approximate Gaussian broadening $\eta$. The left column shows results for $h_x = 0$ ($\eta\approx0.13$), and the right column for $h_x = 0.3$ ($\eta\approx0.14$). Top row: $T=0$. Central row: $T=0.5$. Bottom row: $T=1$.}
 \label{fig: xxz_stag}
\end{figure}

\paragraph{XXZ chain in a staggered field.}

Quasi-one-dimensional spin systems such as Cu-PM \cite{Zvygain_2004, Zvyagin_2011} and copper benzoate \cite{Dender_1996, Asano_2003} possess alternating crystal axes giving rise to nearest-neighbor Dzyaloshinskii-Moriya (DM) interactions or an alternating $g$-tensor. In the presence of a uniform magnetic field $h_z$, both generate an effective staggered field $h_x$ perpendicular to the direction of $h_z$~\cite{Oshikawa_1997, Affleck_1999}. Motivated by these observations, we consider the isotropic Heisenberg model with antiferromagnetic exchange coupling $J=1$ in a staggered transverse field,
\begin{align}
  H_{\text{stag}} = \sum_{i=1}^{L-1} \mathbf{S}_i\cdot\mathbf{S}_{i+1}  + h_z \sum_{i=1}^{L} S^z_i  + h_x \sum_{i=1}^{L}(-1)^i S^x_i.
\end{align}
We focus on the central region of the magnetization curve and in the following keep $h_z=1$ fixed. We
analyze the effect of these symmetry and integrability breaking interactions on the longitudinal spin
structure factor $S_{zz}(k,\omega)$ for $h_x=0.3$ by comparing to results for systems without DM
interactions, i.e., $h_x=0$. Figure~\ref{fig: xxz_stag} shows our results for a system with $L=50$ and OBCs at temperatures $T=0$,~$0.5$, and $1$. We keep $m=120$ states and expand up to order $N=1500$ for $T>0$. At $T=0$, $N=750$ is sufficient to reach the same resolution, since the full spectral range of the Hamiltonian only makes up half of the bandwidth of the Liouvillian. The method addresses each $k$ value individually. Note that the high frequency resolution obtained with the Chebyshev MPS approach enables us to resolve interesting features of the finite-temperature spectral functions which are difficult to see with other methods.

First, we discuss the case $h_x=0$ displayed in the left column of Fig.~\ref{fig: xxz_stag}. The result at $T=0$ obtained via a Chebyshev expansion without the doubled system is shown in Fig.~\ref{fig: xxz_stag}(a). As can be seen, the numerical results agree well with the analytical boundaries for the spin-wave continua from the Bethe ansatz \cite{Mueller_1981}. Well-converged finite-size effects (FSEs) can be resolved for $L=50$. In the lower continuum, the oscillations are FSEs and decay in amplitude towards higher frequencies, most prominently at $k\approx3\pi/4$. Note the tiny peak just above the lower boundary of the lower continuum at $k\approx3\pi/8$ moving to higher frequencies with increasing $k$ and a similar branch visible in the upper spin-wave continuum which may be physical features. The high intensity for $k,\omega \to 0$ occurs due to spin conservation. For our choice of $h_z$, $S_{zz}(k,\omega)$ is gapless at $k=0$ and at $k\approx3\pi/4$. This is representative for a $T=0$ Luttinger liquid (LL) with Fermi momentum $2k_F \approx 3\pi/4$ \cite{Giamarchi}. It is now interesting to see how the LL changes by increasing the temperature. For example, in Refs.~\onlinecite{Feiguin_2010, Bonnes_2012}, the `melting' of a LL for a $t$-$J$ chain with Kondo impurities and for SU($N$) symmetric Hubbard systems, respectively, has been investigated numerically by considering spectral functions at finite (effective) temperatures. Reference~\onlinecite{Bonnes_2012} found the peak at $2 k_F$ to move with temperature to $k = \pi$ at $T \sim 0.2$ in their units of energy. 
  
Here, we find using ED that already for very low temperatures $T<0.1$ the peak at $2k_F$ is significantly broadened, and at temperatures of the order $T \lesssim 0.5$ becomes strongly suppressed. This is seen in Fig.~\ref{fig: xxz_stag}(c), where for $T=0.5$ the signal at $k\approx 3\pi/4$ is replaced by a broad distribution around $k=\pi$. However, the QMC results of Ref.~\onlinecite{Grossjohann_2009} indicate that at $T=0.25$ a feature in the vicinity of $2k_F$ remains visible. It appears interesting to study the evolution of this peak as a function of $T$, which we leave for future investigations. As can be seen in Fig.~\ref{fig: xxz_stag}(e), further increasing the temperature to $T=1$ does not significantly alter the picture. Note that the FSEs observed at $T=0$ are not visible at $T=0.5$ and~$T=1$ due to temperature broadening while the resolution remains the same.

We now turn to the effect of a staggered field of magnitude $h_x=0.3$ on $S_{zz}(k,\omega)$. Comparing Figs.~\ref{fig: xxz_stag}(a) and \ref{fig: xxz_stag}(b), we identify the opening of a field-induced gap at $T=0$ and the formation of a well-defined band.
This is in agreement with the expectations from adding a DM term to the Heisenberg Hamiltonian since it causes the opening of gaps \cite{Oshikawa_1997, Dender_1996} and a mixing of the longitudinal and transverse components of correlation functions, causing the formation of the observed band.  
Interestingly, increasing the temperature from $T=0$ to $T=0.5$ does not significantly alter the results: A redistribution of the weights is obtained and the signals are smeared out, but in contrast to the $h_x=0$ case the qualitative features persist. Further increasing the temperature to $T=1$ leads to a stronger redistribution of the weights and eventually the band disappears.

\begin{figure}[t]
\centering
 \includegraphics[width=0.9\columnwidth]{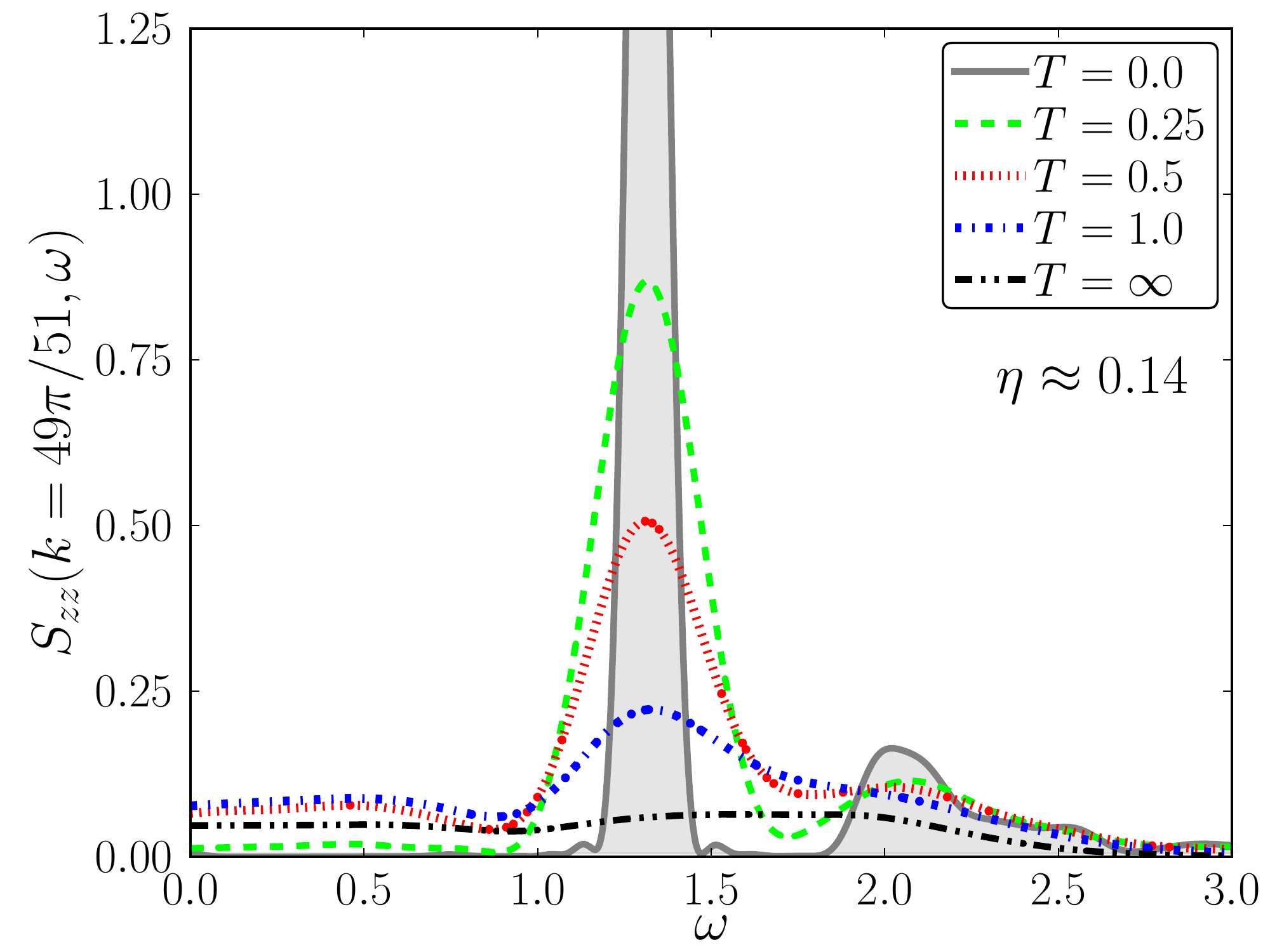}%
 \caption{(Color online) MPS results for the frequency dependence of
 $S_{zz}(k,\omega)$ at $k=49\pi/51$ of an $L=50$ Heisenberg chain in a uniform magnetic field $h_z=1$ and an additional staggered field $h_x=0.3$ at  temperatures $T=0,\,0.25,\,0.5,\,1$, and~$\infty$.}
 \label{fig: xxz_L50_k49}
\end{figure}

These features are studied in more detail in Fig.\ \ref{fig: xxz_L50_k49}, which shows the temperature dependence of the longitudinal spin structure factor for $h_x=0.3$ at $k=49\pi/51$. At higher temperatures $T \gtrsim 0.5$, the peak at $\omega\approx1.3$ shows thermal broadening. Note the filling in of spectral weight into the gap at small frequencies when increasing the temperature, which is due to scattering. The signal corresponding to the excitation at high energies starts to disappear for temperatures of the order of $T=0.5$. Further  increasing the temperature from $T=0.5$ to $T=1$ also significantly broadens the peak at $\omega \approx 1.3$, which represents the band visible at $T=0$. Finally, at infinite temperatures, the curve is rather flat with only small features up to $\omega \approx 2$, when it starts to decay to zero. 

\paragraph{Conclusions and perspectives.}
We presented an efficient and very accurate approach to compute finite-temperature spectral functions of strongly correlated quantum systems directly in the frequency domain by using a Liouville space formulation. We implemented this via a Chebyshev expansion in a DMRG framework, and show additional results from an alternative CFE implementation in the Supplemental Material, demonstrating the flexibility of our approach. The high resolution allowed us to observe the disappearance of the $T=0$ Luttinger liquid upon increasing temperature. 
In contrast, considering the effect of spin-orbit coupling leads to an opening of a field-induced gap and the formation of a band which both remain stable over a wide temperature range. While we focused on proof-of-principle calculations, we expect that the frequency resolution can be further increased by a factor of ten by optimizing the MPS-based Chebyshev expansion \cite{Holzner_2011}. Together with the high flexibility of MPS methods, the Liouville approach will allow for an unbiased and efficient treatment for a variety of systems and to directly compare to experimental results at finite temperatures, as obtained in neutron scattering, electron spin resonance, transport experiments, or  more recently in the context of ultracold gases \cite{Jin_Nature2008}.

\acknowledgments
We acknowledge helpful discussions with Lars Bonnes, Christoph Karrasch, Thomas K\"ohler, and Alexei Kolezhuk as well as computer support by the GWDG and the GoeGrid project. A.H. and A.C.T. thank the Helmholtz Gemeinschaft for financial support via the Virtual Institute ``New states of matter and their excitations'' (Project 6). T.P. and S.R.M. acknowledge support by the CRC SFB 1073 (Project B03).

\bibliography{References}

\end{document}


\title{Matrix product state formulation of frequency-space dynamics at finite temperatures: \textit{Supplemental Material}}

\author{Alexander C. Tiegel}
\affiliation{Institut f\"ur Theoretische Physik, Georg-August-Universit\"at G\"ottingen, 37077 G\"ottingen, Germany}

\author{Salvatore R. Manmana}
\affiliation{Institut f\"ur Theoretische Physik, Georg-August-Universit\"at G\"ottingen, 37077 G\"ottingen, Germany}

\author{Thomas Pruschke}
\affiliation{Institut f\"ur Theoretische Physik, Georg-August-Universit\"at G\"ottingen, 37077 G\"ottingen, Germany}

\author{Andreas Honecker}
\affiliation{Institut f\"ur Theoretische Physik, Georg-August-Universit\"at G\"ottingen, 37077 G\"ottingen, Germany}
\affiliation{Fakult\"at f\"ur Mathematik und Informatik, Georg-August-Universit\"at G\"ottingen, 37073 G\"ottingen, Germany}

\date{\today}                 

\maketitle


 \subsection{Continued fraction expansion}
 This Supplemental Material is intended to demonstrate the flexibility of the Liouville space formulation described in the main text. We envisage that a variety of methods can be applied within this framework.  
In the following, we present results obtained by a continued fraction expansion (CFE) \cite{Haydock_1972, Gagliano_1987, Dagotto_1994, Hallberg_1995} of the Green's function generated by the Lanczos algorithm \cite{Lanczos_1950,Cullum_2002} for the Liouville operator. In this case, the finite-temperature spectral function from Eq.~(2) in the main text is approximated by
 \begin{align}
  I_A(k,\omega) = - \frac{1}{\pi} \, \text{Im} \, \frac{ \left\langle \Psi_T | A^\dagger A | \Psi_T \right\rangle }{z - a_0  - \frac{b_1^2}{z-a_1 - \frac{b_2^2}{z - a_2 - \ldots}} },
 \end{align}
with the coefficients 
$a_i = \frac{\left\langle u_i \left| \mathcal L \right| u_i \right\rangle}{\left\langle u_i | u_i \right\rangle}$ 
and $b_i^2 = \frac{\left\langle u_i | u_i \right\rangle}{\left\langle u_{i-1} |  u_{i-1} \right\rangle}$ of the Lanczos recursion $| u_{i+1} \rangle = \mathcal L | u_i \rangle - a_i | u_i \rangle - b_i^2 | u_{i-1} \rangle$.  
For its evaluation, we use a matrix product state implementation of the Lanczos algorithm, whose $T=0$ version was introduced in Refs.~\onlinecite{Dargel_2011, Dargel_2012}. The starting vector for the Lanczos recursion is $|u_0\rangle = A\,| \Psi_T \rangle / \sqrt{\langle \Psi_T | A^\dagger A | \Psi_T \rangle}$.
  
  \begin{figure}
  \centering
    \includegraphics[width=0.93\columnwidth]{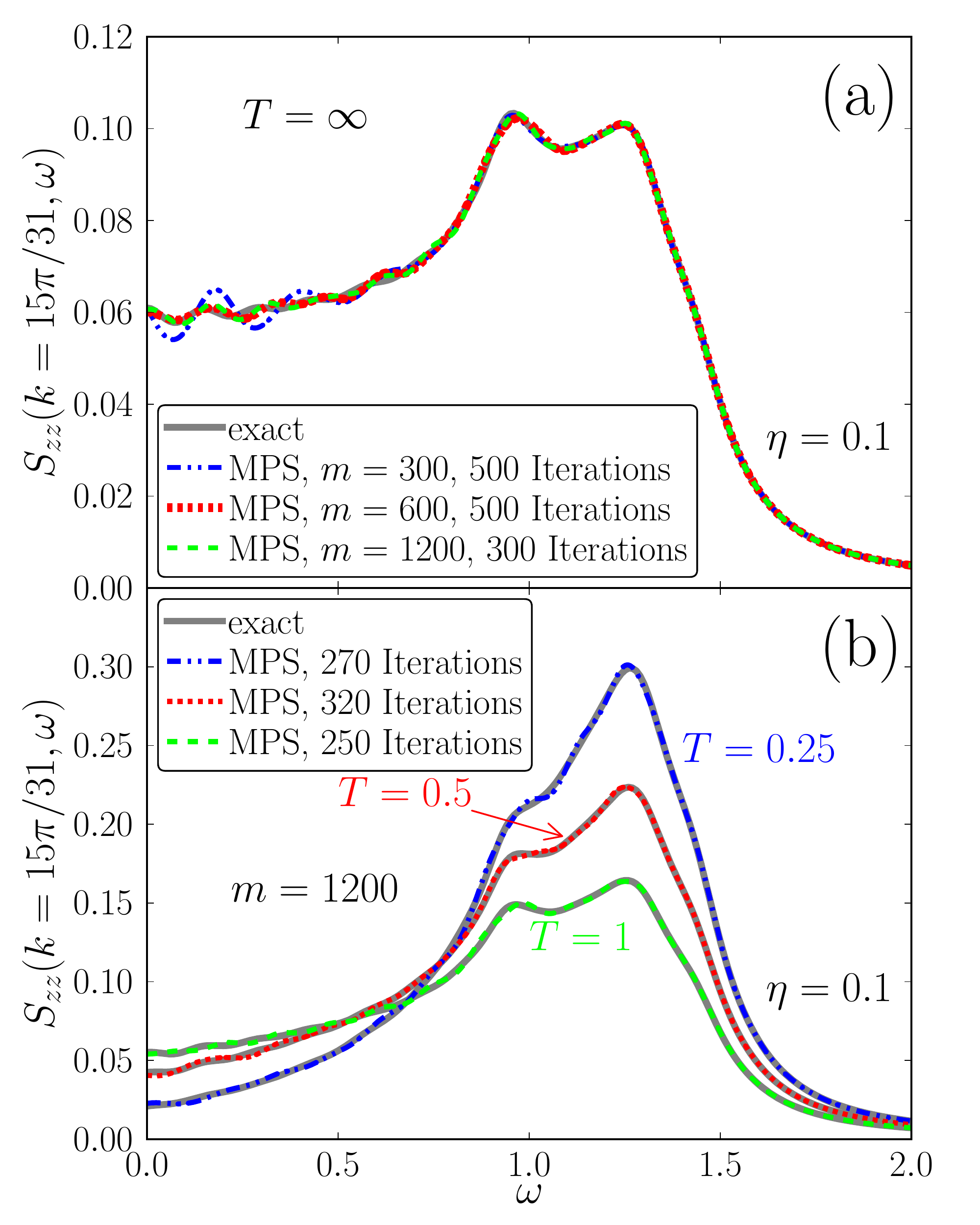} 
  \caption{(Color online) Finite-temperature DMRG calculations using a CFE for the dynamic spin structure factor
  $S_{zz}(k=15\pi/31,\omega)$ of an open $XX$ chain with $L=30$ compared to the exact solution. (a) At
  $T=\infty$ the accuracy is studied for different numbers of kept DMRG states $m$. (b) Temperature dependence of $S_{zz}(k=15\pi/31,\omega)$ compared to the exact solutions for $T=1,\,\,0.5$, and~$0.25$ (solid lines).}
\label{fig: xy_chain_comparison_cfe}
\end{figure}

\paragraph{Comparison to exact results for the $XX$ model.}

In Fig.\,\ref{fig: xy_chain_comparison_cfe} we compare the exact result for the longitudinal spin structure factor of an $XX$ chain with $L=30$ to the one obtained by a CFE. 
The exact results are obtained as described in the main text, except that we adopt a Lorentzian  broadening of $\eta=0.1$ on account of the CFE. We terminate the continued fraction once the spectral function is visibly converged for the chosen value of $\eta$. The agreement of the CFE results with the exact solutions is good as the relative deviations are always $\lesssim 10~\%$ (see below for more details). Comparing the CFE results for $L=30$ to the Chebyshev expansions for $L=50$ shown in Fig.\,1 of the main text, we find that the number of kept DMRG states $m$ for a Chebyshev expansion can be chosen considerably smaller than for a CFE: When devising the Chebyshev expansion, visible deviations cannot be resolved in this plot for as small a value as $m=100$, whereas for the CFE in Fig.~\ref{fig: xy_chain_comparison_cfe} we need to keep a substantially larger number $m \gtrsim 300$ to achieve a comparable accuracy at a still lower resolution. Moreover, the convergence properties of the Chebyshev expansion at finite temperatures also improve over those of the Lanczos algorithm. This becomes particularly clear in Fig.~\ref{fig: xy_chain_comparison_cfe}(a) which shows our CFE results for the longitudinal spin structure factor $S_{zz}(k=15\pi/31,\omega)$ at infinite temperature. Here, the absolute deviation of the CFE from the exact result ($\lesssim 0.005$ for $m=300$ and $\lesssim 0.002$ otherwise) is largest in the center of the spectrum, i.e., at small frequency. This can be explained by the fact that the extremal eigenvalues at the edges of the spectrum converge best in the Lanczos algorithm.

In  Fig.~\ref{fig: xy_chain_comparison_cfe}(b) the temperature dependence of $S_{zz}(k=15\pi/31,\omega)$ for $m=1200$ is depicted. Again, the DMRG results fit the exact curves well down to temperatures as small as $T=0.25$. A detailed analysis reveals that the maximum relative deviation for $m=1200$ is $\sim3.5~\%$ for $T=1$, $\sim6~\%$ for $T=0.5$ and $\sim8.5~\%$ for $T=0.25$ (the absolute deviations are $\lesssim 0.004$ in all cases). Thus, these errors in the spectral functions are mostly due to the CFE and can be lowered by devising a Chebyshev expansion.

\paragraph{$XXZ$ chain in a staggered field.}

\begin{figure}[h] 
\centering
 \includegraphics[width=0.98\columnwidth]{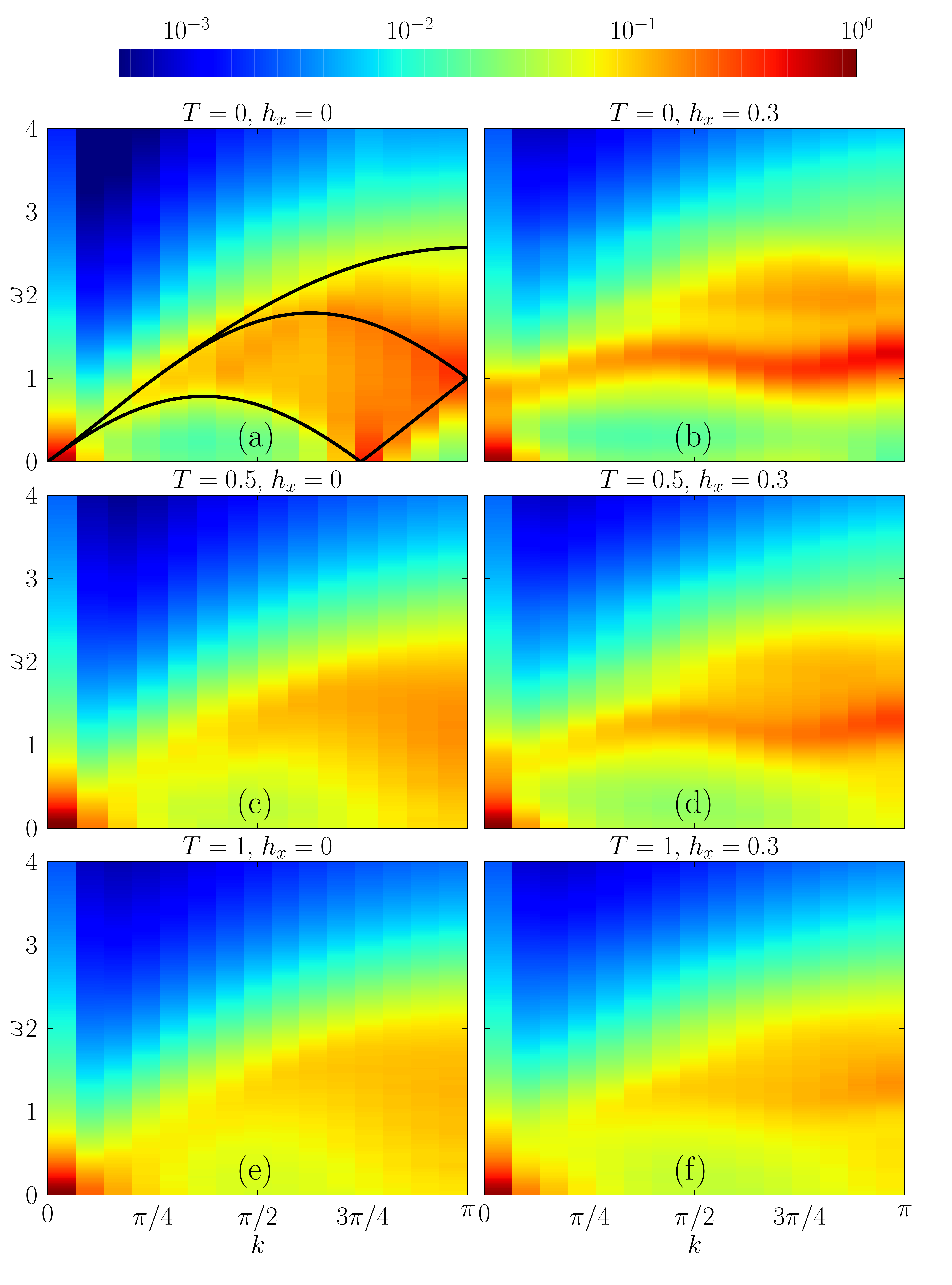} 
 \caption{(Color online) Longitudinal spin structure factor $S_{zz}(k,\omega)$ of a Heisenberg chain with $L=30$ in a magnetic field $h_z = 1$ obtained with the CFE approach ($\eta=0.2$). The left column shows results for $h_x = 0$, and the right column for $h_x = 0.3$. Top row: $T=0$. Central row: $T=0.5$. Bottom row: $T=1$.}
 \label{fig: xxz_stag_cfe}
\end{figure}

In Fig.~\ref{fig: xxz_stag_cfe}, we furthermore show the momentum-resolved spin structure factor of the isotropic $XXZ$ chain in a uniform magnetic field $h_z$ with and without a staggered magnetic field $h_x$ at temperatures $T=0$,~$0.5$, and $1$ obtained via a CFE. All panels of this figure correspond to those of Fig.~2 in the main text featuring the Chebyshev results. All main features of the dynamic spin structure factor are also resolved by the CFE. However, in the Lanczos algorithm we kept $m=500$ DMRG states for a chain of $L=30$ and perform at least 300 iterations. This yields a well converged CFE with a Lorentzian broadening of $\eta=0.2$. In contrast, the Chebyshev expansion only requires $m=120$ states for $L=50$ at an even enhanced resolution.
 
\newpage
The high quality of the CFE data demonstrates the flexibility of the Liouvillian formulation. It remains an interesting question for future investigations to study the efficiency of further approaches applied in this framework, such as correction-vector methods \cite{Jeckelmann_2002, Weichselbaum_2009}. 

\bibliography{References}